# CRYSPNet: Crystal Structure Predictions via Neural Network

Haotong Liang[1*], Valentin Stanev[1,2,3*], A. Gilad Kusne[4,1,2], Ichiro Takeuchi[1,2]

[1]Department of Materials Science and Engineering, University of Maryland, College Park, MD 20742, USA

[2]Maryland Quantum Materials Center, University of Maryland, College Park, MD 20742, USA

[3]Joint Quantum Institute, University of Maryland, College Park, MD 20742, USA

[4]National Institute of Standards and Technology, Gaithersburg, MD 20899, USA

*These authors contributed equally to this work

Structure is the most basic and important property of crystalline solids; it determines directly or indirectly most materials characteristics. However, predicting crystal structure of solids remains a formidable and not fully solved problem. Standard theoretical tools for this task are computationally expensive and at times inaccurate. Here we present an alternative approach utilizing machine learning for crystal structure prediction. We developed a tool—Crystal Structure Prediction Network (CRYSPNet)— that can predict the Bravais lattice, space group, and lattice parameters of an inorganic material based only on its chemical composition. CRYSPNet consists of a series of neural network models, using as inputs predictors aggregating the properties of the elements constituting the compound. It was trained and validated on more than 100,000 entries from the Inorganic Crystal Structure Database. The tool demonstrates robust predictive capability and outperforms alternative strategies by a large margin. Made available to the public (at https://github.com/AuroraLHT/cryspnet), it can be used both as an independent prediction engine or as a method to generate candidate structures for further computational and/or experimental validation.

## I. INTRODUCTION

Finding new materials with desired properties remains one of the grand challenges in science. However, the current demand for novel materials far exceeds the capabilities of traditional approaches combining trial-and-error and serendipitous discoveries. Only a combination of more efficient experimental strategies[1] with computational methods which can reliably and quickly predict properties of materials *in silico*[2] can fully address this crucial need.

Structure is one of the most fundamental properties of crystalline solids; it determines directly or indirectly the majority of materials characteristics. Crystal structure is also the starting point for the first-principles computational tools used to calculate many materials properties of practical interest[2]. However, predicting the crystal structure itself remains a formidable and not fully solved problem[3,4]. Most current methods for this task rely on calculating the energy of a large set of candidate structures (typically generated randomly or using some similarity function) to find the best global solution. Despite the great advances in optimization methods and the dramatic increase in the available computing power, this approach remains arduous and computationally expensive. The problems become severe as the size and the complexity of the studied systems rises, driving an exponential increase in the dimensionality of the search space and the number of possible solutions. Moreover, first-principles methods are typically built on uncontrolled approximations which sometimes lead to poor accuracy of the formation energy calculations.

Machine learning (ML) methods have recently emerged as a new, powerful approach in the study of materials[5-10]. In particular, ML provides an alternative to using first-principles calculations.

Instead of solving complex quantum-mechanical problems directly, ML methods can make predictions based on correlations found in measured or calculated data. These correlations, in turn, are learned through statistical and probabilistic methods[5,11]. Although suffering from a fundamental limitation, namely inability to predict outcomes (e.g., structures) not in the training data, ML methods have several significant advantages. They are less susceptible to human biases and erroneous assumptions, especially when trained on experimental data. Use of ML can circumvent some of the limitations of even the most sophisticated *ab initio* methods, such as difficulties calculating properties for finite temperatures and modeling compositionally-modified (through substitution) non-stoichiometric compounds. Once an ML model is trained, it typically provides an extremely fast and inexpensive means to generate predictions.

Despite these advantages, ML has not been used extensively for crystal structure predictions. Furthermore, the works that utilized ML methods for this task tended to focus on particular materials groups[12-16]. This approach leads to specialized models trained on limited data and with restricted applicability. Such models clearly cannot be used to predict the likely structure of an arbitrary hypothetical composition. To goal of our work is to address this gap and to explore the possibility of using ML methods for general crystal structure prediction. We present here a tool named CRYSPNet, designed to predict the Bravais lattice, space group, and lattice parameters of an inorganic material solely based on its chemical composition. To develop the tool we utilized more than 100,000 entries from the Inorganic Crystal Structure Database (ICSD)[17]. The access to such large dataset allowed us to extensively train and validate the ML models. For input, the tool relies on aggregate predictors based on the properties of the elements constituting the compound. Similar approach has been successfully used to predict other materials properties, from band gap energies[18,19] to superconducting and ferromagnetic critical temperatures[20,21], and metallic glass formation abilities[22].

CRYSPNet (abbreviated from Crystal Structure Prediction Network) utilizes Neural Networks (NN) – the class of ML models behind many of the recent breakthroughs in AI applications in science and technology[23]. The biggest advantage of NNs is their ability to perform feature selection and training simultaneously, obviating the need for manual creation and selection of predictors. In condensed matter physics and materials science, NNs have been successfully applied to a variety of tasks, ranging from analysis of Monte Carlo simulations[24,25], microscopy images[26,27] and x-ray diffraction data[28,29], to molecular design[30,31] and knowledge extraction from published data[32,33]. Yet, to the best of our knowledge, this is the first work to apply these methods directly for crystal structure prediction (although they have been used for related tasks such as classifying X-ray diffraction patterns[34] and predicting possible compositions forming a given structure[35]).

CRYSPNet demonstrates good predictive capability and unconditionally outperforms trivial strategies such as random or mode selection. In addition to using the entire dataset based on ICSD, we developed models for two subsets, namely, oxides and metallic alloys. Surprisingly, the NN models trained on the latter—relatively small—dataset containing only around 16,000 entries show much better overall performance compared to the models for oxide dataset contains almost 56,000 compositions. They are able to confidently predict the Bravais lattice, space group, and lattice parameters with great accuracy. The models trained on the entire and oxide datasets are generally less reliable, but also decisively outperform trivial strategies, demonstrating their ability to learn from the available data.

This tool is freely available and can be downloaded at https://github.com/AuroraLHT/cryspnet. It is easy to use, and can be utilized both as a stand-alone engine for crystal structure predictions or as a rational and data-informed way to generate candidate structures for further computational and experimental validation. Thus, it can be used in conjunction with more sophisticated (but also more demanding) approaches such as first-principles computations.

## II. DATA AND MODELS

### A. ICSD dataset

ICSD is one of the largest collections of crystal structures of inorganic solids[17]. Currently, it has more than 210,000 entries, containing structural information about solids ranging from pure elements to extremely complex compounds with more than ten constituent elements. To construct a dataset of crystal structures, we have extracted the chemical formulas, Bravais lattices, space groups, lattice parameters, and other relevant information from 181,362 unique ICSD entries.

After obtaining this information, a manual data cleaning was performed in order to make the notation uniform. It has to be noted that the notation used in "symmetry_cell_setting" field of the CIFs[36] (the basic data structure of the database, containing information for a single entry in ICSD) is neither lattice system nor crystal system. The trigonal label is used interchangeably with hexagonal and rhombohedral labels for the same space group. We mapped the symmetry cell field of all the trigonal compounds (total of 18,068 entries) to the corresponding lattice system and Bravais lattice based on their space group. Since rhombohedral lattices can be represented in both hexagonal and rhombohedral unit cells, we converted all rhombohedral entries with a hexagonal unit cell to a rhombohedral unit cell (see Appendix A). Thus, all materials were categorized in 14 Bravais lattices, with "P", "I", "C", and "F" as short-hand notation for primitive, body-centered, base-centered with unique $c$-axis, and face-centered.

The existence of polymorphs leads to multiple entries for compounds with the same chemical formula. This is problematic for our approach: since it relies only on elemental features (see below), it is highly desirable for each unique chemical formula to correspond to a single crystal structure. To remove the duplication caused by the polymorphs, a simple algorithm was used to determine the most likely stable phase at ambient conditions (prevalent for ICSD entries). We selected entries by comparing the abundances of compounds with the same space group, as well as the ground state formation energy from the Materials Project database. (The details of this procedure are provided in Appendix B.) After removal of the polymorphs the dataset contains a total of 110,813 unique composition-structure pairs. Henceforth, we will refer to this as the Master Dataset.

The Master Dataset allows straightforward survey of 110,000 entries in ICSD providing "panoramic views" of most known inorganic compounds. Fig. 1(a) shows the distribution of solids in the Master Dataset by the number of constituent elements. As can be seen here, the binary, ternary, quaternary, and quinary materials comprise about 95% of the dataset, while compounds with more elements are very rarely observed. (Since the number of possible combinations grows exponentially with the number of elements, this sparseness underlines the fact that the space of complex multi-component materials is still almost unexplored.) Fig. 1(b) shows the abundance of the top 12 elements. Oxygen, the most abundant element, is present in more than 50% of the entries, indicating the dataset has a strong preference for oxide compounds. The data can also be used to study the history of how compounds were discovered and entered into ICSD (see Appendix C).

It is well known that compounds in ICSD are highly unevenly distributed over the 14 Bravais lattices and the 230 space groups[37]. The abundance of each Bravais lattice is shown in Fig. 1(c) and as can be seen, about 40% of the materials are in the cubic (F), orthorhombic (P), and hexagonal (P) lattices. In contrast, the number of orthorhombic (F) or orthorhombic (I) entries is relatively small. As shown in Figure 1(d), the relative abundance of different Bravais lattices also depends on the number of elements – triclinic and monoclinic entries are rare in binary phases but become more common as the number of elements increases, underscoring the correlation between stoichiometric complexity and low-symmetry structures. Large imbalance is also observed in the distribution of space groups, where $P2_1/c(14)$, $Pnma(62)$, $Fd\bar{3}m(227)$, $Fm\bar{3}m(225)$, $P\bar{1}(2)$, $P6_3/mmc(194)$, $I4/mmm(139)$, $C2/c(15)$, $C2/m(12)$, and $R\bar{3}m(166)$, the ten most abundant space groups, encompass almost half (47%) of the dataset (see Figure 2(a)).

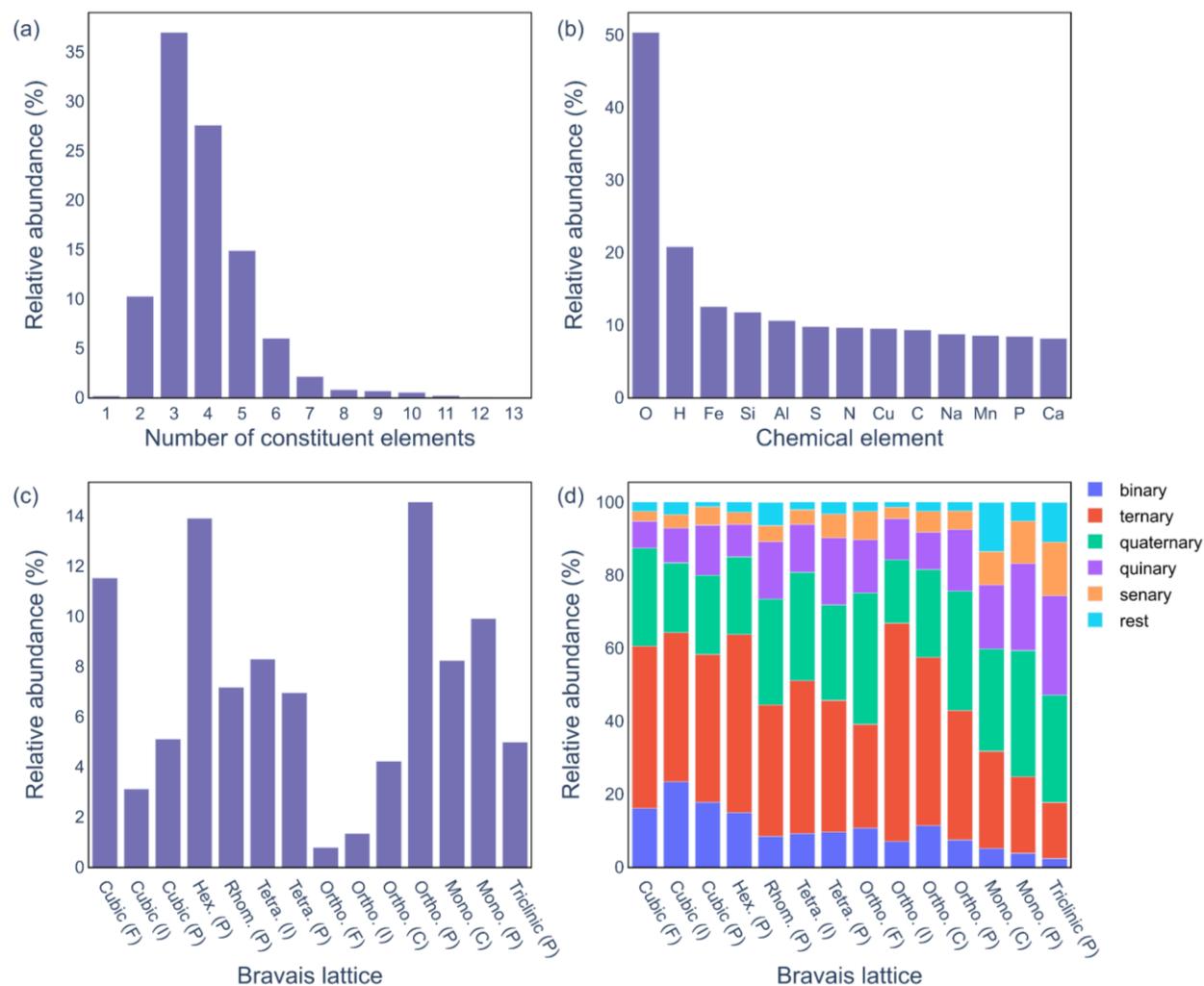

**Fig. 1.** Survey of compounds in ICSD: (a) Relative abundance of compounds with varying number of constituent elements; (b) Relative abundance of the top 13 most ubiquitous elements in the dataset; (c) The percentage of each Bravais lattice in the dataset; (d) The distribution of systems with varying number of elements with the same Bravais lattice class. Compounds with more than six elements are combined in a single group. The abbreviations are "Hex." for hexagonal, "Rhom." for rhombohedral, "Tetra." for tetragonal, "Ortho." for orthogonal, "Mono." for monoclinic. "P", "I", "C", and "F" denote primitive, body-centered, base-centered with unique c-axis, and face-centered systems, respectively.

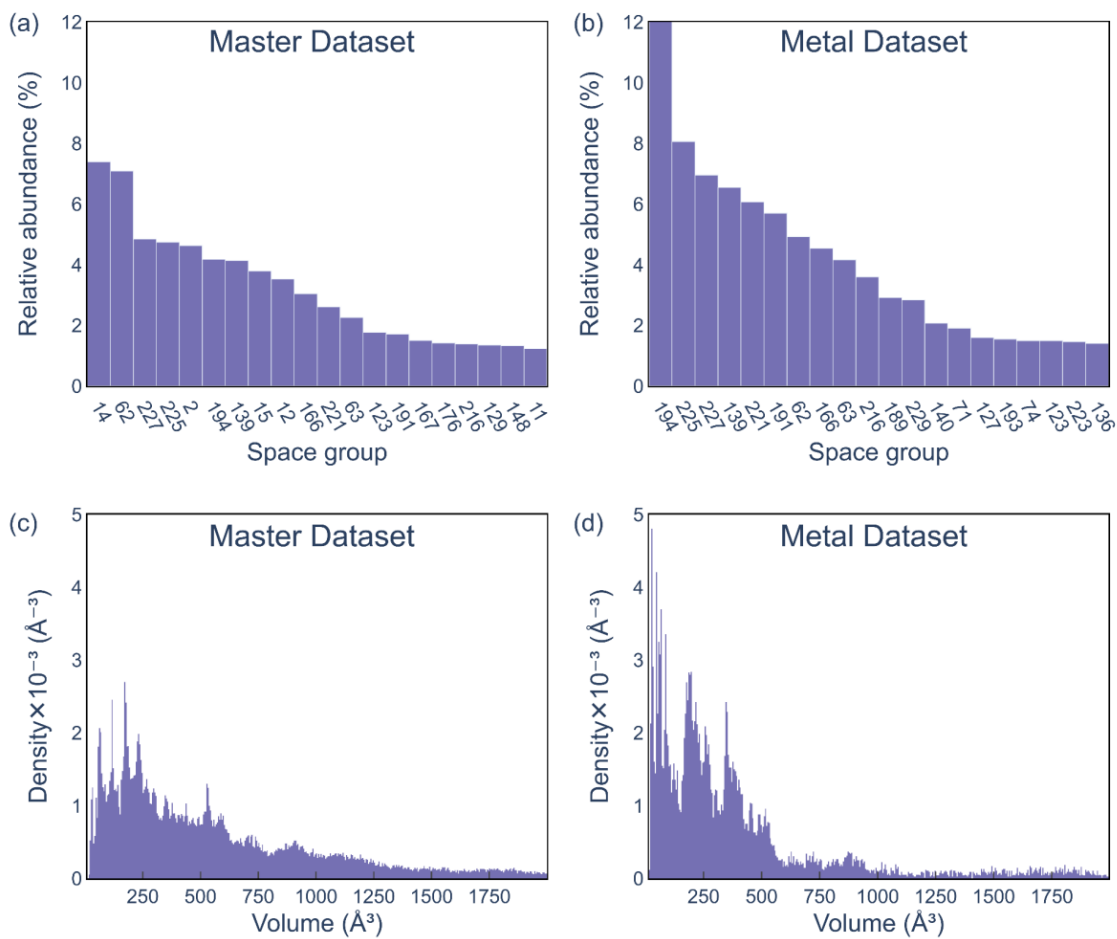

**Fig. 2.** Distribution of datasets by space group and volume: (a) The relative abundance of the top-20 most common space group in the Master Dataset.; (b) The relative abundance of the top-20 most common space group in the Metal dataset; (c) The histogram of unit cell volume in the Master Dataset; (d) The histogram of unit cell volume in the Metal Dataset;

To be able to analyze the data in more detail, we extracted two subsets from the Master Dataset: one with only oxide compounds (Oxide Dataset, total of 55,770 compositions); and another set with compounds formed by elements in alkali, alkaline earth, transition, post-transition, lanthanoid, and actinoid metal series (Metal Dataset, with 16,127 entries). Analyses on these subsets show that the Metal Dataset has a distinct distribution of space groups and unit cell volumes, different from the Master Dataset. As can be seen in Fig. 2(b), the list of most abundant space groups of the Metal Dataset contains such high symmetry space groups like $P6_3/\mathrm{mmc}(194)$, $\mathrm{Fm}\bar{3}\mathrm{m}(225)$, and $\mathrm{Fd}\bar{3}\mathrm{m}(227)$. Meanwhile, the Master Dataset is dominated by low symmetry space groups (Fig. 2(a)). We also found that unit cell volume of the Metal Dataset is generally smaller and more tightly concentrated in the range of up to 500 Å³ (shown in fig. 2(c)-(d)). As mentioned before, more than half of ICSD is composed of oxide compounds, and thus, the symmetry and lattice patterns for Oxide Dataset are quite similar to the patterns of the Master Dataset. As described in Section III, all three datasets are used to train separate neural networks to predict Bravais lattices.

### B. Predictors

To generate predictors for the ML modeling we used Matminer library[38]. It is an open-source Python library designed to automate several steps in the process of data mining properties of materials. We utilized its predefined feature generation methods that transform chemical composition into materials predictors. 132 Magpie predictors[18] were employed as a starting point. They consist of the minimum, maximum, mode, weighted average (referred to as "mean"), and weighted average deviation (denoted by "avg dev") over a specific compositional/elemental property. (The mean and the average deviation are calculated by $\bar{f} = \sum x_i f_i$ and $\hat{f} = \sum x_i |f_i - \bar{f}|$, respectively. $f_i$ denotes the value of the property for an element $i$, and $x_i$ is the mole fraction of this element in any given compound.) These properties include number of elements, positions in the periodic table, covalent radius, electronegativity, number of valence electrons in each orbit, as well as space group, specific volume, band gap energy, and magnetic moment of the ground state ($T = 0$ K) elemental structure. We removed all features that are based on the mode. These only change when the majority component changes, which introduces discontinuities in the predictor space and makes them problematic for predicting entries with atomic substitutions. In addition to the Magpie set, some extra features were added, expanding the total feature set size to 228. A list of some important elemental properties used to generate predictors for the models is shown in Table 1.

Table 1. Materials predictors used to generate the datasets from compositional information. Magpie predictor set is based on calculating the mean, average deviation, minimum, and maximum of the elemental properties (weighted by the fraction of each element in the composition). Computing the additional predictors differs from that for the Magpie set. For more details, the reader can consult the Matminer documentation[38].

| Magpie Predictors | Additional Predictors |
|---|---|
| Atomic Number | Stoichiometry p-norm (p=0,2,3,5,7) |
| Mendeleev Number | Elemental Fraction |
| Atomic Weight | Fraction of Electrons in each Orbital |
| Melting Temperature | Band Center |
| Periodic Table Row & Column | Ion Property (possible to form ionic compound, ionic charge) |
| Covalent Radius | |
| The number of Valence e- in each Orbital (s, p, d, f, total) | |
| The number of unfilled e- in each orbital (s, p, d, f, total) | |
| Ground State Band Gap Energy | |
| Ground State Magnetic Moment | |

### C. Machine learning models

We combined several Multi-Layer Perceptron (MLP) models to create a machine learning tool, CRYSPNet, able to predict the Bravais lattice, space group, and the lattice parameter of a material based solely on the elemental predictors described in the previous section. MLP is a NN architecture that stacks densely connected layers, combining a series of non-linear activation functions and optional regularization steps. ReLU ($f(x) = \max(0, x)$) function is used as activation function on the inner layers. We adopt Dropout and BatchNorm as two effective regularization methods used to prevent overfitting[39,40]. (More details about the architecture of the MLP models are given in Appendix D.) We employed two distinct MLP architectures: one to predict the Bravais lattice and the space group, and another one for the lattice parameters

The last layer of the MLP yields the model predictions. For the classification problems (i.e., predicting the Bravais lattices and the space group labels) we utilize the softmax function, which gives the predicted probability for the $i$-th compound to be in the $l$-th class by $y_{l,i} = e^{x_{l,i}}/\sum_{l=1}^{K} e^{x_{l,i}}$, where $K$ is the total number of classes (e.g., 14 Bravais lattices) and $x$ is the output from the previous layer. These models minimize the cross-entropy loss $\mathcal{L}_{CE}$ (based on negative log likelihood for correct prediction):

$$\mathcal{L}_{CE} = -\frac{1}{N}\sum_{i=1}^{N}\sum_{l=1}^{K} t_i(l) \log(y_{l,i})$$

where $N$ is the total number of data points, and $t_i(l)$ is the indicator variable for the true label of the $i$-th compounds ($t_i(l)$ is 1 for $l = l_{true}$ and 0 for all other $l$).

For the lattice parameters predictions (a regression problem), the final layer output uses a sigmoid function and a scaling factor (applied to confine the predictions into a physically meaningful range). The loss function is the Log Mean Square Error (Log-MSE) $\mathcal{L}_{\text{Log MSE}}$:

$$\mathcal{L}_{\text{Log MSE}} = \frac{1}{N}\frac{1}{D}\sum_{i=1}^{N}\sum_{j=1}^{D}\left[\log\left(\frac{Y_{ij}}{t_{ij}}\right)\right]^2,$$

where $N$ is again the total number of data points, $y_{ij}$ is the prediction for the $j$-th parameter from the set of $D$ symmetry group-based parameters for the $i$-th compound (e.g. $D = 1$ for primitive cubic but $= 6$ for primitive triclinic), and $t_{ij}$ is the ground truth value for the $i$-th prediction. The benefit of using Log-MSE as a loss function is that it measures the relative rather than absolute errors (ratio vs. Euclidean distance). This is important since entries with larger lattice parameters tend to have larger prediction errors. Using absolute errors would have biased the model towards trying to predict better the larger lattices parameters, to the detriment of all others.

The schematic diagram of the CRYSPNet's workflow is shown in Fig. 3. Note that the Bravias lattice model precedes the other two—space group and lattice parameters—NN models, which are independent of each other. Since the lattice system determines the number of independent lattice parameters, prediction of the Bravais lattice has to come before the corresponding model for lattice parameters. For example, a model for a (likely) cubic compound only needs to predict a single lattice parameter $a$, whereas a model for triclinic compounds has to predict six: $a$, $b$, $c$, $\alpha$, $\beta$, and $\gamma$. To capture this, 14 distinct models were trained on materials that share the same Bravais lattice. Similarly, the possible space groups are also constrained by the Bravais lattice, and thus 14 distinct models for space group prediction were trained.

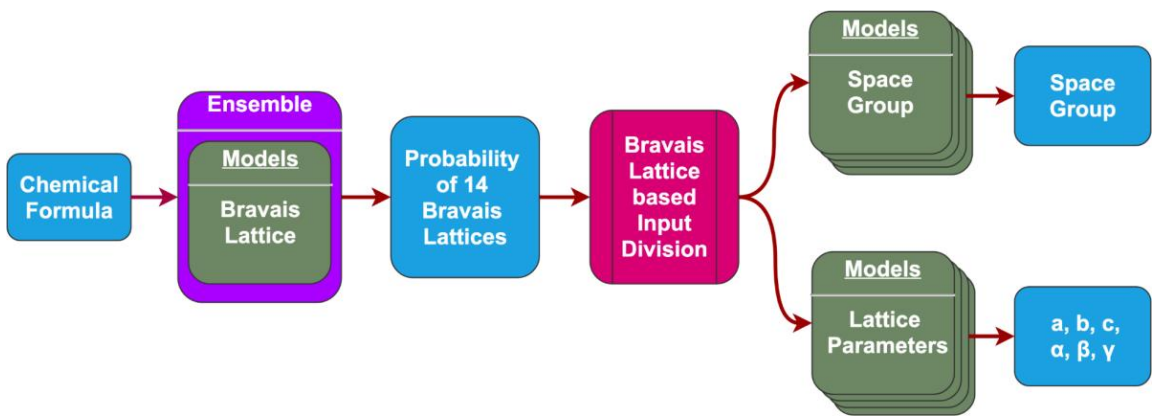

**Fig. 3.** The workflow of CRYSPNet. The input requires the user to provide one or more valid chemical formulas to the tool, which transforms it to a matrix of materials predictors. This matrix is then first used to predict the Bravais lattice of the materials by an ensemble of neural network (NN) models. Depending on the predicted Bravais lattice type, specific NN models are used to predict the lattice parameters and space groups of the materials.

### D. Metrics

Predicting the Bravais lattices and the space groups is a multi-label classification problem. To evaluate the performance of the models, accuracy and top-k accuracy were used as metrics. (The top-k accuracy measures the proportion of correct predictions in the $k$ classes with highest predicted probability). Top-k accuracy with $k \geq 2$ is often used for multi-label classification tasks: the canonical (top-1) accuracy can be too strict a measure, especially if the probabilities for several of the top ranked classes are close and all of them are of interest. For instance, the top three likeliest Bravais lattices can be used to initialize parallel first-principles calculations.

Predicting the lattice parameters is generally a multivariate regression problem (since the number of lattice parameters is larger than one for all except the cubic lattices). Moreover, some of the variables have different physical meaning (length vs angle). To simplify evaluating and comparing different models we use as a main regression metric the coefficient of determination $R^2$, defined as follows:

$$R^2 = 1 - \frac{\sum_{i=1}^{N}(y_i - t_i)^2}{\sum_{i=1}^{N}(\bar{t} - t_i)^2},$$

where $y_i$ and $t_i$ denote the predicted and true values of the lattice parameter of $i$-th compound, and $\bar{t}$ is the average value over the entire dataset. $R^2$ values close to 1 and 0 suggest the model have excellent or no predictive power, respectively.

To avoid overfitting and obtain unbiased estimates of these metrics, we use separate subsets to train and benchmark the models. All datasets were divided into training and validation sets using a separation probability of 0.1. The model parameters were optimized using the training set, while the metrics were calculated on the validation set.

### E. Interpreting the models

Interpretability of ML models is of great importance, especially when they are applied to scientific problems. Unfortunately, NN models are notoriously difficult to interpret directly, due to their multi-layer nature (creating a series of non-linear transformations of the input predictors). To gain an insight into the inner workings of the models, some indirect methods are typically required.

Permutation Importance is a commonly used and relatively simple approach to estimate the relative importance of each of the predictors. It quantifies the significance of a predictor by measuring the validation error with this predictor's values being randomly shuffled, while keeping everything else fixed. To obtain a robust measure of the importance, $K$-times validation is performed with the same model for each predictor (the dataset is randomly split in $K$ subsets, and all possible $K$-1 groups are used to train the model, while the remaining points are used as a validation set). If a predictor is important for the model, its shuffling should lead to a high validation error. Thus, the higher the error, the more important the feature is. We calculate the permutation importance of the features for all the models we created.

The Permutation Importance method provides some insights into how the input predictors are used in the decision-making process. To obtain a more complete understanding of the way the model actually encodes knowledge in its native space, it is useful to study the activation of the hidden units. For example, in computer vision, visualizing the weights of a hidden layer is used to discover regions and features the model considers important[41]. In materials science, it was demonstrated that the activations of a model studying chemical space encoded the structure of the Periodic Table[35].

One challenge in processing the activations of a large NN model is the so-called curse of dimensionality – since the average distances scale with the number of dimensions, points become very sparsely distributed in high-dimensional spaces. Thus, conventional clustering algorithms (such as K-Means Clustering, Gaussian Mixtures, and DBSCAN) do not perform well when applied directly on such high-dimensional data. Instead, dimensionality reduction techniques (e.g., Principal Component Analysis (PCA), t-distributed Stochastic Neighbor Embedding (t-SNE), and Multi-Dimensional Scaling (MDS)) are often used to first project the points to a lower-dimensional space and then cluster the projections.

We utilized a similar approach in order to understand how well the models can distinguish different materials groups. t-SNE algorithm is used to project the activations on a two-dimensional space; these projected points are then clustered by DBSCAN algorithm (detailed description is provided in Appendix E). As we demonstrate in the next section, these clusters indeed represent meaningful grouping of materials, and can be analyzed to understand the model's internal representation of compounds. The clustering also presents an alternative approach for searching materials with structural and chemical similarity, which can yield candidates for future exploration.

## III. RESULTS AND DISCUSSION
### A. Predicting the Bravais lattice

CRYSPNet consists of several distinct components (see Fig. 3). The first is a Bravais lattice prediction module. We trained and tested this module on Master, Oxide, and Metal Datasets, leading to three separate Bravais lattice prediction models, namely, the Master Model, the Oxide Model, and the Metal Model. The performance of these models is shown in Table 2. Despite the fact it is trained with the smallest dataset, the Metal Model reached the highest accuracy of about 70% and top-2 accuracy value of ≈84%, whereas the Oxide Model has the lowest accuracy of 54% and top-2 accuracy of 71%. The performance of the Master Model is similar to that of the Oxide Model, with accuracy and top-2 accuracy of 55% and 71%, respectively. To put these numbers in perspective, they have to be compared with some available alternatives. A random selection of the Bravais lattice from the empirical distribution of observed compounds will lead to an average accuracy of 16%, 23%, and 17% for Oxide, Metal and Master Dataset, respectively. Always selecting the mode (the most popular class) yields a similar accuracy. Thus, the NN models have clearly learned from the data and are able to greatly outperform the other (trivial) strategies.

In Fig. 4 we show the confusion matrix for each model, together with the true and predicted distributions of the numbers of compounds of each class in the validation sets. As can be seen from the figure, in some cases the poor performance (low numbers on the diagonal of the confusion matrix) of the models is clearly connected with the scarcity of data points. It implies the severe problems with class imbalance are harming the performance of the models. This affects especially the low symmetry compounds in the Metal Dataset (Fig. 4(b)), which are typically hard to synthesize in a laboratory or difficult to find in nature. On the other hand, the matrices for Oxide and Master Datasets show the models do not perform too well on the monoclinic (C, P), orthorhombic (C, F, P), and triclinic (P) entries, regardless of their abundancies. However, further analysis shows the second highest probability prediction of these entries typically match the target Bravais lattice. The accuracy and top-2 accuracy of these classes are plotted in Fig. 4(d), and it can be seen there is a large improvement (20% - 30%) from top-1 to top-2 accuracy of triclinic (P), orthorhombic (P, C), and monoclinic (P, C) entries. Thus, the low accuracy for these classes is mostly due to the inability of the model to select correctly between the top two predictions. Low performance in orthorhombic (C, F) is probably explained separately by insufficient training data. As more data become available in the future, a performance boost in these two classes is expected.

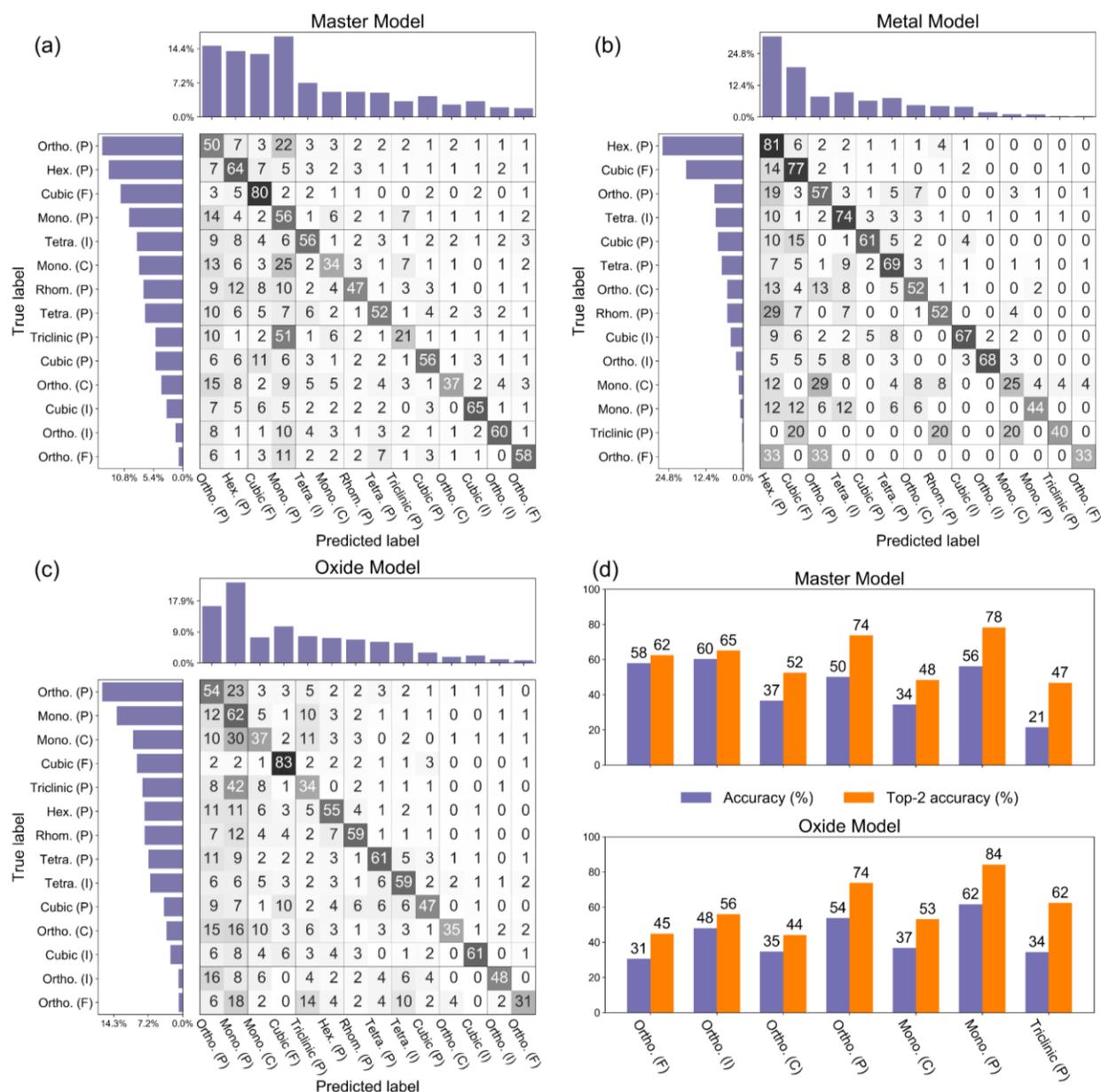

**Fig. 4**. Performance of the model predicting the Bravais lattices: (a) Confusion matrix of the Master Model; (b) Confusion matrix of the Metal Model; (c) Confusion matrix of the Oxide Model. The grey shade corresponds to the relative density of the true labels (also provided as percentages). The numbers on the diagonal shows the prediction accuracy of each Bravais lattice. Histograms on the left and on the top represent the distribution of the true and predicted labels, respectively. (d) The (top-1) accuracy and top-2 accuracy of the Master Model (top) and Oxide Model (bottom). The abbreviations are "Hex." for hexagonal, "Rhom." for rhombohedral, "Tetra." for tetragonal, "Ortho." for orthogonal, "Mono." for monoclinic. "P", "I", "C", and "F" denote primitive, body-centered, base-centered with unique c-axis, and face-centered system, respectively.

**Table 2.** The accuracy and top-2 accuracy of models predicting the Bravais lattice trained on the Master, Metal and Oxide datasets. The results are the average and the standard deviation over 15 models trained for each dataset.

| Dataset | accuracy model | accuracy random | top-2 accuracy model |
|---|---|---|---|
| Master | 54.2±.4 % | 16.5 % | 70.6±.4 % |
| Metal  | 69.5±.9 % | 23.0 % | 84.2±.7 % |
| Oxide  | 55.0±.8 % | 17.1 % | 71.1±.4 % |

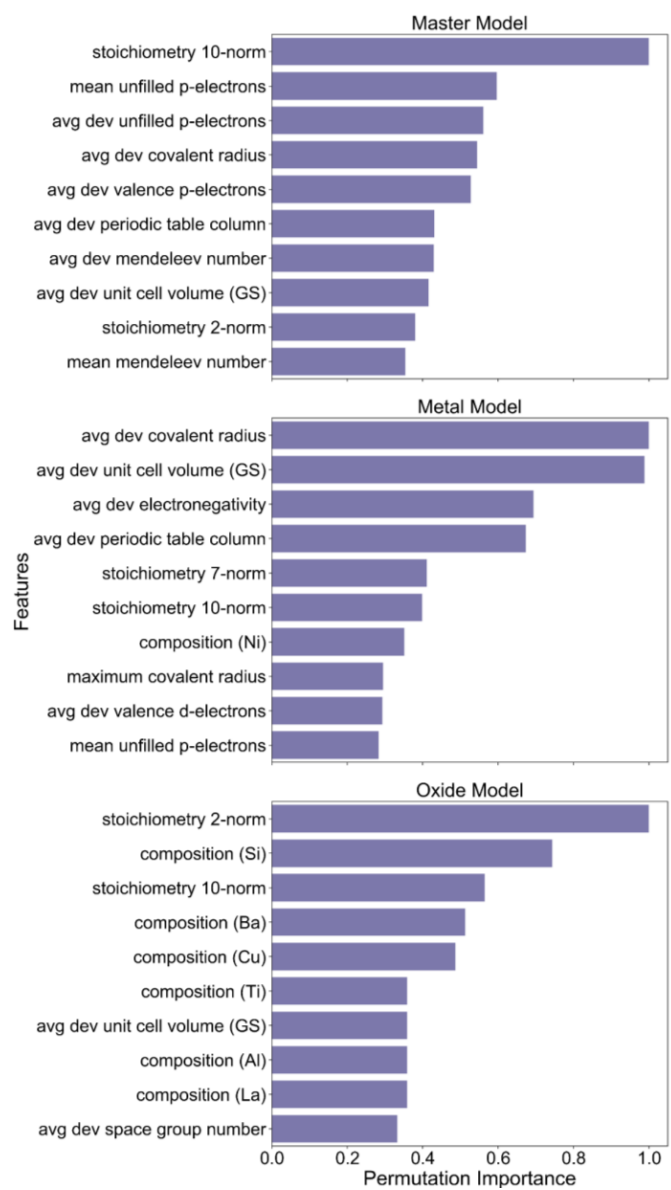

**Fig. 5.** Permutation Importance of the models for Bravais lattices prediction trained on Master, Oxide, and Metal Datasets. The permutation importance of each model is normalized by the feature with the maximum validation error.

To measure the contribution of each feature, the permutation importance method was applied for all three models (shown in Fig. 5). For the Metal Model the top four features (*avg dev covalent radius*, *avg dev unit cell volume at ground state*, *avg dev electronegativity* and *avg dev periodic table column*) hold a large portion of the overall feature importance. The presence of predictors based on the covalent radii and the ground state volumes indicates that the atomic sizes of each element greatly affect the crystal structure of the metal compound, hardly a surprising finding. The relative importance of *avg dev periodic table column*, *mean periodic table column* and those *valence electron number* features show the significance of the elemental electronic structure. The Master Model's list of most important features contains many of the same features as the Metal Model, again indicating the significant of these predictors. By contrast, none of these features are in the top four important features for the Oxide Model. It is relying much more on the compositional features, which are, unfortunately, less interpretable. It suggests, however, that different branches of the model have to be created for particular oxide materials.

Despite the models being trained on very different datasets, some common features are present in all three. One such common point is the use of the *stoichiometry p-norm* features. The lower *p-norm (p=2) can* capture the spread of the composition, while the higher *p-norm (p=7, 10)* is more sensitive to the largest fraction of the composition. By using these features the models are capable to learn that materials with similar stoichiometries are likely to share the same crystal structures (e.g., $ABX_3$ materials usually have perovskite structure, while $X_3Y_2(SiO_4)_3$ materials usually have garnet structure).

### B. Predicting the space group

After predicting the Bravais lattice class, we move on to predicting the compound's space group. To restrict the possible space group range, we trained 14 independent models, one for each Bravais lattice class. The Master Dataset is first grouped into 14 subsets based on the Bravais lattice, each of which is then used as the input for the corresponding model (see Fig. 3). The architecture of the NN is similar to the model in the previous section. During the training process, the space group and Bravais lattice models are trained independently. It should be noted that the space group models are thus based on the assumption that the Bravais lattices predictions are correct.

The cross-validated accuracies and top-3 accuracies of all 14 models are shown in Table 3 (estimated under the assumption of an accurate Bravais lattice determination). Note that since the triclinic system has only two (and highly imbalanced at that) space groups, both its accuracy and top-3 accuracy are automatically very high, and we do not include it in the subsequent discussion. Somewhat surprisingly, models for cubic type lattice perform significantly better than those for non-cubic type lattices, with accuracy in the 87-91% range. Among the remaining 10 models for non-cubic types, the accuracy is typically in the 70-80% range. Again, these results should be compared with the random selection and mode selection strategies. Applying random selection yields the accuracy range from 13-58%, where the highest accuracy is for monoclinic (C) systems. Meanwhile, selecting the most popular space group yield the accuracy range from 25-74%, where the highest accuracy is for monoclinic (P) systems, and the second largest accuracy (59%) is for orthorhombic (I) compounds. This high accuracy for the second trivial method of selecting space groups demonstrate the very imbalanced distribution of data mentioned in section 1. In all the cases (except for the triclinic system) our model outperforms both trivial strategies, leading to a dramatic increase in the accuracy (in some cases with more that 40-50%), again demonstrating the ability of the models to extract meaningful information from the data.

**Table 3.** The cross-validated accuracy and top-3 accuracy of models predicting the space group trained for the 14 Bravais lattices. The cross-validated results were obtained after averaging the results of 15 independent training runs for each Bravais lattice class. The 100% top-3 accuracy in the triclinic row is a trivial result—there are only two space groups in this Bravais lattice. The abbreviations are "Hex." for hexagonal, "Rhom." for rhombohedral, "Tetra." for tetragonal, "Ortho." for orthogonal, "Mono." for monoclinic. "P", "I", "C", and "F" denote primitive, body-centered, base-centered with unique c-axis, and face-centered system, respectively.

| Bravais lattice | accuracy model | accuracy random | top-3 accuracy model |
|---|---|---|---|
| Cubic (F) | 90.7±0.6 % | 36.5 % | 98.8±0.2% |
| Cubic (I) | 87.1±1.7 % | 17.6 % | 95.1±1.2% |
| Cubic (P) | 87.3±1.6 % | 30.4 % | 95.8±0.1% |
| Hex. (P) | 74.8±0.9 % | 13.3 % | 87.9±0.6% |
| Rhom. (P) | 81.7±1.4 % | 26.6 % | 94.5±1.0% |
| Tetra. (I) | 81.8±1.1 % | 29.1 % | 92.7±1.0% |
| Tetra. (P) | 78.2±1.4 % | 13.6 % | 86.9±1.0% |
| Ortho. (F) | 72.7±4.7 % | 24.8 % | 93.0±0.3% |
| Ortho. (I) | 78.0±3.1 % | 40.6 % | 91.9±2.6% |
| Ortho. (C) | 72.2±1.9 % | 33.0 % | 90.8±0.8% |
| Ortho. (P) | 63.8±1.4 % | 25.6 % | 81.0±1.0% |
| Mono. (C) | 74.0±1.1 % | 40.1 % | 94.0±0.7% |
| Mono. (P) | 79.4±1.1 % | 58.0 % | 81.0±1.0% |
| Triclinic (P) | 94.7±0.9% | 87.1 % | 100% |

### C. Predicting the lattice parameters

Similar to the space group prediction model, we use 14 subsets, grouped by Bravais lattice, to create and train 14 separate models for predicting the lattice parameters (see Fig. 3). The validation error (Log-MSE) of each model, as well as the $R^2$ for each relevant lattice parameter are shown in Table 4. As can be seen there, the validation error increases as the degree of symmetry of the Bravais lattice decreases. The model trained on cubic (P) entries has the lowest Log-MSE value of 0.12, while the validation error for the triclinic one is six times higher. Since the number of lattice parameters also increases six-fold, there is clear deterioration of the models' performance as the number of parameters to be predicted increases. Interestingly, for the triclinic type the predictions for the angles $\alpha$, $\beta$, and $\gamma$ tend to be much worse than the predictions for $a, b$, and $c$ parameters (see also Fig. A3 in Appendix F). Similar trend appears in the $\beta$ predictions for the monoclinic type. We believe this peculiar property of the model at least partially originates in the crystallographic rules and conventions for representing lattices. For monoclinic unit cells, there are multiple rules that define the selection of basic vectors, which can easily confuse the model. For example, in monoclinic (P), if there is a glide operation, the $c$-axis is selected to parallel to that translation, whereas $c$-axis is to be selected to satisfy $a \leq c$ if there is no glide operation. However, the convention prefers a fully reduced mesh unit cell where $a$ and $c$ are the two shortest transition vectors and thus the interaxial angle β < 120°. The restriction in $c$-axis might lead to a different selection of $a$-axis (and β) which might result in a different setting than fully reduced cell. In our dataset, 11% and 20% of the entries in Monoclinic (P, C) are not in fully reduced setting. Selection becomes even more complicated in triclinic system. The reduced cell are classified as type *I* (α < 90°, β < 90°, and γ < 90°) and *type II* (α ≥ 90°, β ≥ 90°, and γ ≥ 90°). We observe both type *I* and type *II* unit cells in our dataset, with other configurations also present. Similar reasoning can also explain the inferior performance of the model on orthorhombic entries. For orthorhombic lattices, c-axis is preferred to be the one that has lower symmetry element than $a$ and $b$ axes. When the symmetry is same for all axes, they are selected so that $a \leq b \leq c$. In our dataset, 23% of the orthorhombic (P, I, F) entries are given in the latter format.

Table 4. The cross-validated Log-MSE for the models predicting the lattice constants for each Bravais lattice. $R^2$ values can be used to assess the performance of the prediction models for each lattice constant. The results present the average and the standard deviation over 15 models trained for each subset. The abbreviations are "Hex." for hexagonal, "Rhom." for rhombohedral, "Tetra." for tetragonal, "Ortho." for orthogonal, "Mono." for monoclinic. "P", "I", "C", and "F" denote primitive, body-centered, base-centered with unique c-axis, and face-centered system, respectively.

| Bravais Lattices | Log MSE | $R^2$ | | | | | |
|---|---|---|---|---|---|---|---|
| | | $a$ | $b$ | $c$ | α | β | γ |
| Cubic (F) | 0.22±.02 | 0.83±.02 | - | - | - | - | - |
| Cubic (I) | 0.16±.06 | 0.80±.06 | - | - | - | - | - |
| Cubic (P) | 0.12±.03 | 0.85±.03 | - | - | - | - | - |
| Hex. (P) | 0.27±.02 | 0.77±.02 | - | 0.61±.04 | - | - | - |
| Rhom. (P) | 0.30±.03 | 0.52±.08 | - | - | 0.71±.04 | - | - |
| Tetra. (I) | 0.21±.02 | 0.73±.03 | - | 0.74±.03 | - | - | - |
| Tetra. (P) | 0.21±.01 | 0.76±.03 | - | 0.64±.09 | - | - | - |
| Ortho. (F) | 0.51±.11 | 0.19±.43 | 0.39±.16 | 0.39±.16 | - | - | - |
| Ortho. (I) | 0.65±.10 | 0.51±.11 | 0.35±.11 | 0.33±.09 | - | - | - |
| Ortho. (C) | 0.48±.04 | 0.39±.06 | 0.43±.09 | 0.64±.04 | - | - | - |
| Ortho. (P) | 0.50±.02 | 0.42±.02 | 0.35±.03 | 0.37±.03 | - | - | - |
| Mono. (C) | 0.58±.05 | 0.33±.07 | 0.46±.10 | 0.32±.04 | - | 0.17±.05 | - |
| Mono. (P) | 0.76±.02 | 0.11±.13 | 0.22±.03 | 0.19±.03 | - | 0.15±.04 | - |
| Triclinic (P) | 0.80±.04 | 0.24±.08 | 0.24±.17 | 0.22±.07 | 0.03±.04 | 0.06±.03 | 0.01±.03 |

D. **Generalization tests**

Even though we have demonstrated the ability of CRYSPNet to predict the Bravais lattices and other structural parameters for randomly selected compounds in ICSD, the ultimate goal is to have a tool which is able to predict the crystal structures of yet-to-be synthesized materials. To further validate this idea, we split the ICSD dataset based on publication date. Specifically, compounds with structures published before 2014 were used to train the model, while the rest of the dataset was used as a holdout test set. We again split the train and the holdout data in Master, Metal, and Oxide subsets, which are used to train and test respective Master, Metal and Oxide models. The accuracy of the Bravais lattice predictions on the holdout test set is ≈48% for the Master model and ≈51% for Oxide model – roughly similar to the accuracy reported above. On the other hand, we observe a significant difference in the performance of the Metal Model. The accuracy on the holdout test set is 43%, which is much lower than the previous validation accuracy of 69% (albeit still significantly better than what can be achieved using trivial prediction methods). To understand this discrepancy, we analyzed the distribution of compounds by the number of elements in each dataset. For the Master and Oxide Datasets, the distributions of the training and holdout sets is closely matched, leading to consistent model results. In contrast, for the Metal Dataset we found that the portion of binary systems decreased significantly, from 38% in the training set to 15% in the holdout set. Conversely, the combined number of ternary and quaternary systems rose from 61% in the training set to 82% of the holdout dataset. This shift in the underlying distributions can explain the deterioration in the Metal Model performance – after all, we are testing the model on a dataset which is significantly different from the set used to train it.

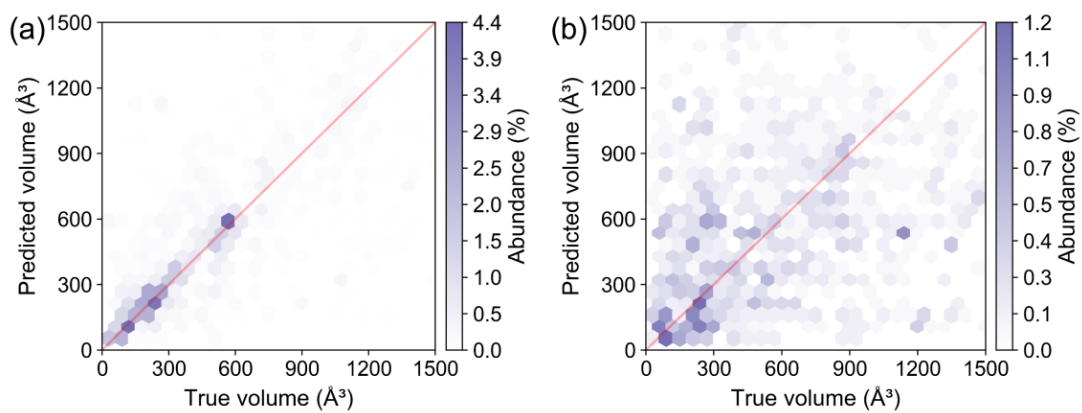

**Fig. 6**. The true unit cell volume vs the unit cell volume computed from the lattice parameters predicted by the models. Entries with correctly classified Bravais lattices are shown in (a), while misclassifies entries are shown in (b).

This result signifies the evolution of studies of materials, with more complex compounds emerging to the forefront of materials research. This gradual shift presents a real challenge to ML methods for predicting crystal structures, as well as other materials properties. Careful tailoring of the training datasets might be required for best performance on relatively new and scarce structures.

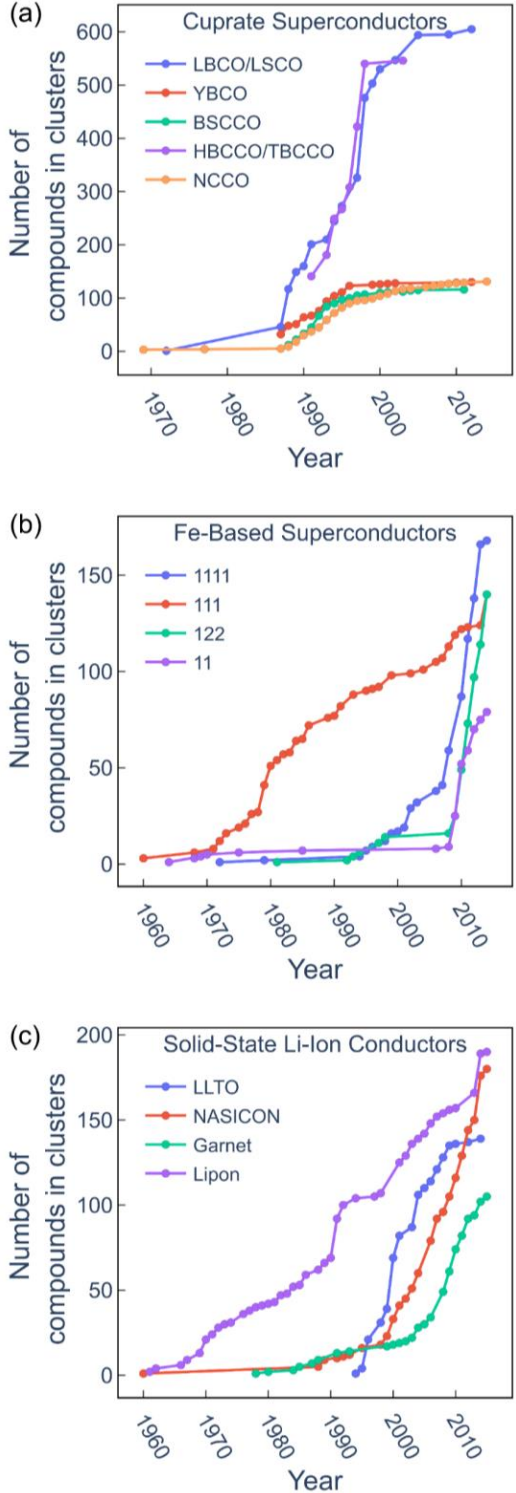

Looking at the general architecture of CRYSPNet (Fig. 3), it is clear that obtaining accurate predictions for the Bravais lattice is crucial for the further performance of the tool. We also investigated how the possible misclassification of the Bravais lattice would affect the subsequence lattice parameter predictions (since each Bravais lattice class has a corresponding lattice parameter model). In Fig. 6 we show the unit cell volume computed from predicted lattice parameter and the experimental determined volume for all materials in the holdout Master Dataset. For the materials with correctly predicted Bravais lattices (Fig. 6(a)) the model leads to a good (although not outstanding) accuracy, with an $R^2$ value of 0.68. The model for misclassified materials (Fig. 6(b)) has an $R^2$ of only 0.04; it is statistically equivalent to simply using the mean lattice parameters as predicted values. Thus—and not very surprisingly—the lattice parameter predictions for materials with incorrectly assigned Bravais lattices are almost meaningless, while predictions for materials with the correct Bravais lattice class can be quite accurate.

### E. Clustering of materials

To gain some insights into the inner workings of CRYSPNet, we use t-SNE to project onto a two-dimensional space the activations of the second hidden layer of the Bravais lattice model trained on the Master Dataset. These projected points are then clustered utilizing a modified version of the DBSCAN algorithm. We first group the projected activations by year, and then for each group perform DBSCAN to form a materials cluster that has similar activations. Time analysis was performed by connecting the clusters that are close together but have different timestamps (see Appendix E for more details). Following this approach, we were able to extract 1534 materials groups from the Master Dataset/Model. To demonstrate that these groups based on projected activations indeed combine related materials, we extracted small samples from clusters that include various cuprate superconductors, iron-based superconductors and Li-ion solid-state electrolytes (see Appendix G).

In Fig. 7 we show the change with time in the number of compounds in these clusters. Their evolution mirrors

**Fig. 7.** Number of compounds as a function of year for several groups that contain interesting materials: (a) high-temperature cuprate superconductors. ("LBCO": $La_{5-x}Ba_xCu_5O_{5(3-y)}$; "LSCO": $La_{2-x}Sr_xCuO_4$; "YBCO": $YBa_2Cu_3O_{7-x}$; "BSCCO": $Bi_2Sr_2CaCu_2O_{8+x}$; "HBCCO": $HgBa_2Ca_2Cu_3O_{8+x}$; "TBCCO": $Tl_2Ba_2Ca_2Cu_3O_{10}$; "NCCO": $Nd_{2-x}Ce_xCuO_4$); (b) high-temperature iron-base superconductors. ("1111": LaOFeP; "111": LiFeAs; "112": $BaFe_2As_2$; "11": FeSe); (c) solid-state lithium-ion conductors. ("LLTO": $Li_{3x}La_{2/3-x}TiO_3$; "NASICON": sodium (Na) super ionic conductor–$LiTi_2(PO_4)_3$; "Garnet": $Li_7La_3Zr_2O_{12}$; "Lipon": lithium phosphorus oxynitride-$Li_2PO_2N$)

some well-known historical developments in their respective fields, further validating the clustering approach. For example, by looking at the number of compounds in different cuprate clusters (Fig. 7(a)) one can clearly see a surge of research activities in the mid-eighties, which in fact corresponds to the discovery of the first cuprate superconductor $La_{5-x}Ba_xCu_5O_{5(3-y)}$ by J. G. Bednorz and K. A. Müller in 1986[42]. The curves indicate that following an initial surge period, activities to identify new compounds belonging to each family tend to plateau after 5-10 years. In a similar manner (Fig. 7(b)), the increase in the numbers of materials in the group containing the iron-based superconductors started naturally after the discovery of the first superconducting system LaOFeP in 2006[43]. In 2008, first reports of iron-arsenic and iron-chalcogenide superconductors further accelerated research[44]. It is interesting to note that the so-called "1111" (e.g. LaOFeP) and "111" (e.g. LiFeAs) pnictides were only discovered to be superconductors in 2006 and 2008, respectively, but their family curves clearly indicate that related compounds from these groups (such as LiCoAs and NaZnAs from the "111" cluster and LaOZnP and LaONiP from the "1111" cluster) had been known for a lot longer. Indeed, it is known that the serendipitous discovery of superconductivity in LaOFeP happened "…in the course of exploration of magnetic semiconductors as an extension of research on transparent p-type semiconductor…"[45] in the La$T_M$OPn ($T_M$ = 3d transition metal, $Pn$ = pnictogen) "1111"-type compounds.

The clusters derived from the NN model can thus be used to visualize and study the evolution of particular materials groups. However, they can potentially have another—more important and innovative—role. The groups created by the clustering are based on similarity/dissimilarity between materials. This similarity combines chemical proximity (as quantified by the input predictors) and structural information (due to the backpropagation optimization of the model), and is an unconventional way to create lists of related materials. These can be then used as stating points in the search for novel compounds exhibiting specific properties (such as superconductivity or high ion conductivity) or other functional properties such as electrocatalysis and thermoelectricity.

## IV. CONCLUSSION

We presented an ML tool, called CRYSPNet, designed to predict the Bravais lattices, space groups and lattice parameters of inorganic solids. CRYSPNet is a combination of several NN models; it was trained on more than 100,000 existing compounds extracted from ICSD. The performance of the tool depends significantly on the exact composition of the training/validation subsets, with the models doing notably better with metallic compounds and high symmetry crystal structures. However, in all cases CRYSPNet significantly outperforms trivial crystal structure prediction strategies.

This work demonstrates that NNs (and, more generally, ML models) provide a viable way to obtain fast, inexpensive, and rational crystal structure predictions given only chemical composition of materials. The relevance of such novel methods is rapidly growing; the ability to confidently predict properties and screen hitherto unexplored materials on a large scale is becoming of paramount importance in materials science and engineering. Reliable and accessible methods for structure predictions should be an integral part of any rational materials design. ML tools can be used for this role, both as stand-alone predictors, or as a first step and a way to constrain the search space of first-principles crystal structure computations.

## ACKNOWLEDGEMENTS

The authors are grateful to Peter Zavalij, Jason Hattrick Simpers, Brian DeCost, and Johnpierre Paglione for valuable discussions and suggestions, and to Stephan Rühl for help with ICSD. This research is supported by AFOSR FA9550-14-10332, ONR N00014-13-1-0635, ONR N00014-15-2-222, and NIST grant # 60NANB19D027.

## APPENDIX A: CONVERSION FROM HEXAGONAL TYPE UNIT CELL TO RHOMBOHEDRAL TYPE UNIT CELL

The equation for converting the lattice parameter is shown below:

$$a_r = b_r = c_r = \sqrt{\frac{a_h^2}{3} + \frac{c_h^2}{9}},$$

$$\alpha_r = \beta_r = \gamma_r = \cos^{-1}\left(\frac{2c_h^2 - 3a_h^2}{2(c_h^2 + 3a_h^2)}\right),$$

where $a_r$, $b_r$, $c_r$, $\alpha_r$, $\beta_r$, and $\gamma_r$ are lattice parameters for rhombohedral unit cell while $a_h$ and $c_h$ are the lattice parameters for hexagonal unit cell.

## APPENDIX B: SELECTION OF ENTRIES

Due to the fact that the ICSD contains multiple entries with the same chemical composition but with different crystal structure, an additional algorithm was needed to determine the most likely stable state at room temperature and ambient pressure (the predominant condition for ICSD entries). An external dataset with 83989 compounds from the Material Project database was used for this purpose. The selection algorithm is based on a score value that combines the abundance in ICSD and the ground state formation entropy ($E_{hull}$, T = 0K) above the convex hull in the Material Project database. The score for formula that coexist in ICSD and Material Project database is computed as follows:

$$\text{Score}(f, s) = \frac{\text{Abundance}(f,\ s)}{E_{hull}(f,\ s) + \alpha}$$

where α (= 0.1 in this paper) is a tunable parameter to balance the formation energy term and abundance count from the ICSD dataset, $s$ is space group, and $f$ is the chemical formula. The $\text{Abundance}(f, s)$ function counts the number of records in ICSD that have the same chemical composition and space group. The $E_{hull}(f, s)$ function finds the lowest formation energy above the convex hull for a given composition and space group. If a given composition only exists in ICSD, the score of its space group is taken as $\text{Abundance}(f, s)$. The most favorable space group is picked from the one with highest $\text{Score}(f, s)$. The most likely structure entry is selected as the one with a unit cell volume closest to the median of unit cell volume of entries in the favorable group.

After utilizing the information from ICSD and Material Project dataset, for each chemical composition we were able to select the entry that has a high abundance in space group value, a low formation entropy, and a reasonable unit cell size. To validate this algorithm, we picked 150 entries and compared their Bravais lattice and space group with their room temperature phases reported in literature. The accuracy for the Bravais lattice is 90%, while the one of space group is 88%, which could be considered an excellent performance for such a simple algorithm.

## APPENDIX C: TIME EVOLUTION OF ICSD

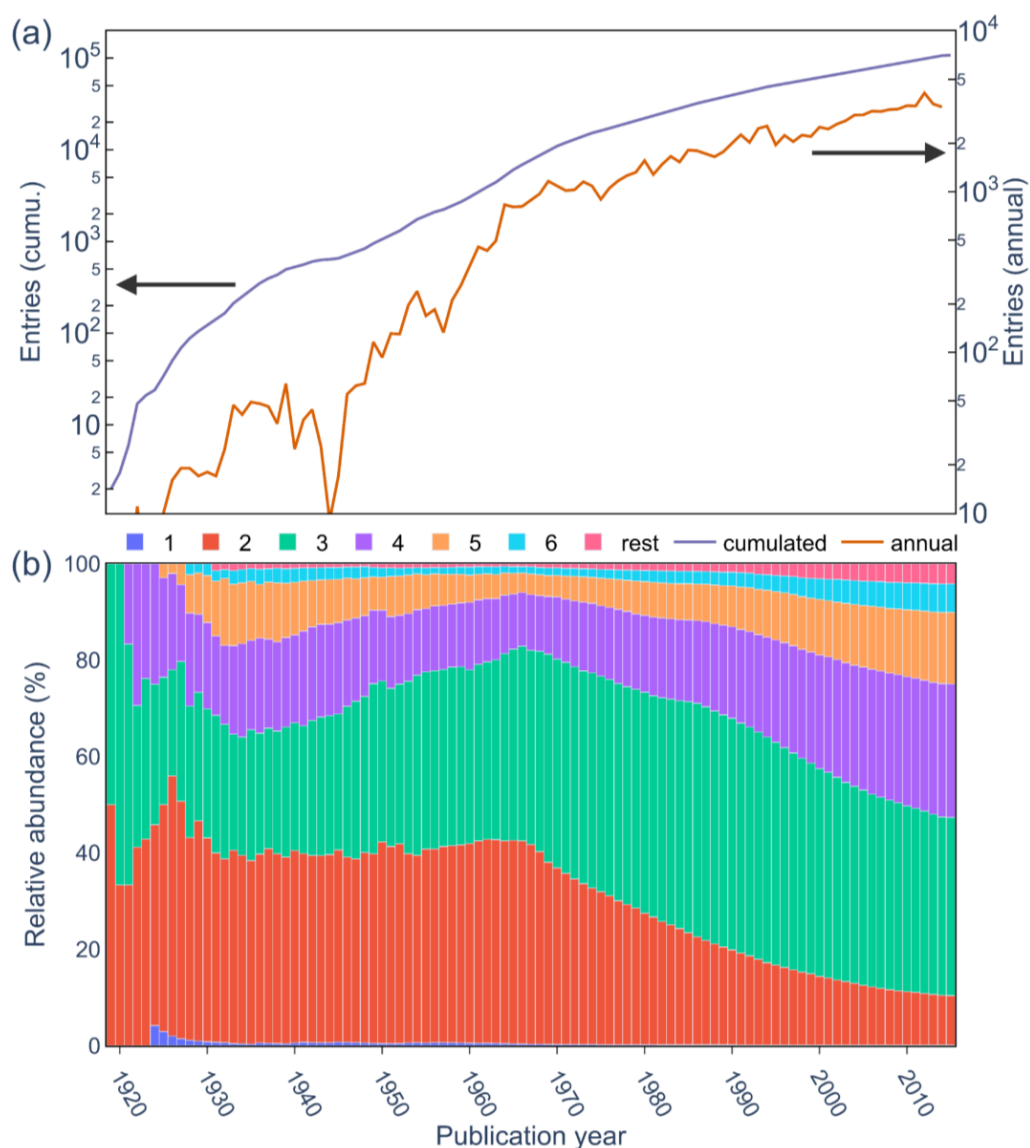

**Fig. A1.** Time evolution of ICSD. (a) Cumulative and annual number of materials entries which were discovered since 1920. (b) Relative abundance of materials grouped by the number of constituent elements and normalized by the total amount of discovered materials as a function of time.

ICSD data can be used to map the evolution of studies of materials. To demonstrate the changes in the focus of materials research over time, we plot in Fig. A1 the numbers of materials entries (both annual and cumulative) in ICSD for each year. The rapid growth in the number of newly discovered materials started early in the last century. In Fig. A1 (b) we show the distribution of materials in ICSD by the compounds with different number of elements. As can be seen, the exploration of materials shifted from binary materials in the early days of 20$^{th}$ century to more complex systems after 1965. This shift signifies two related developments in the study of materials: the number of novel binaries was (nearly) exhausted, and the advent of x-ray detectors enabled the analysis of complex structures beyond binaries.

## APPENDIX D: NEURAL NETWORKS ARCHITECTURE AND TRAINING

The Fastai library (https://github.com/fastai/fastai) was used as the basic toolbox to design the NN models. This library provides a consistent and convenient API for crating and training NN models while remains highly customizable. It integrates multiple best practices accumulated from a large set of diverse Deep Learning applications. One of its tools is a learning rate finder, which helps determining the highest stable learning rate; this allows to reduce the training time and to achieve better generalization (see Appendix H for details). Cyclic Learning Rate policy (cycle length=10 epochs) is also used, both to accelerate and improve the stability of the training process (see Appendix I). A large batch size of 256 and a large dropout probability of 0.1 were employed to enforce better regularization on the models. Each model was trained until the validation error converged. Adam Optimization technique ($\beta_1 = 0.9$, $\beta_2 = 0.999$) was used to backpropagate gradients and update weights of the neural network[47].

The size of each hidden layer of the NN model was determined by hyperparameter optimization using Tree-structured Parzen Estimator Approach (TPE) and Median Pruner implemented by Optuna[48]. The exact architecture for the models is shown on Figure A2.

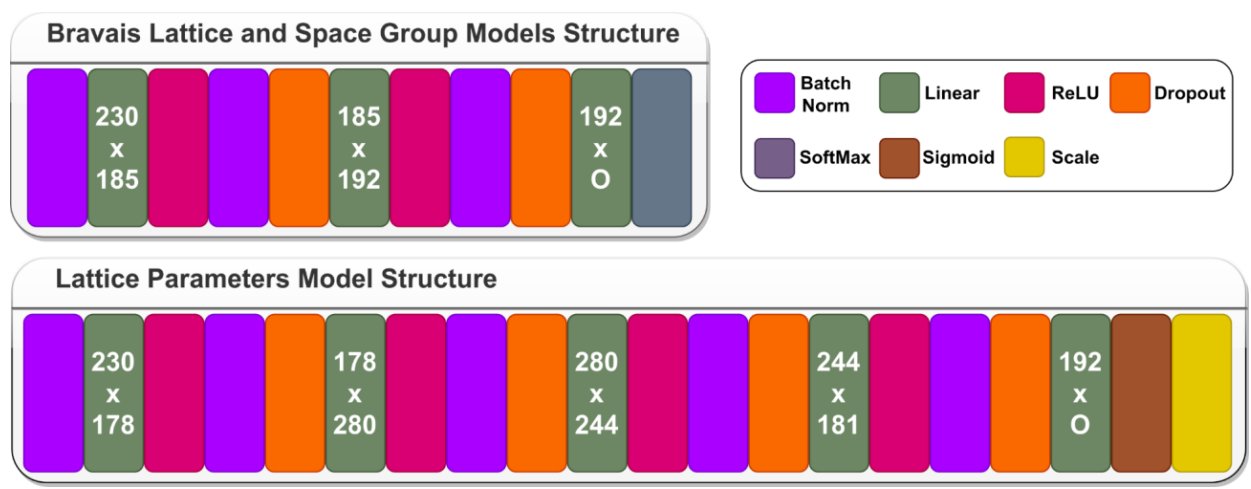

**Fig. A2.** The general architecture of the Neural Network (NN) model for Bravais lattice, space group, and lattice parameters. Each color-coded block represents a layer inside the NN. Model for all application share the same structure except the last output module. The softmax function is used for Bravais lattice and space group predictions. The sigmoid function and a scaling factor are used for lattice parameters prediction.

High variances and high biases were observed in the predictions of a single neural network. This was likely caused by the selection of the training data and/or the stochastic nature of the weight initialization and dropout function. Ensemble Learning with Vertical Voting was deployed to produce stable, low variance, reproducible predictions[49]. Vertical voting utilizes multiple models, trained on different datasets that are randomly sampled from the original data; the outputs of these models are averaged in order to produce a low variance prediction. The equation for vertical voting is

$$P(x \in X) = \frac{1}{M}\sum_{i=0}^{M} P(x \in X | \theta_i),$$

where $X$ is the set of Bravais lattices or space groups, $\theta_i$ represent the parameters of the $i$-th model, and $M$ is the number of models.

To mitigate the problem of class-imbalance, oversampling was applied to the minority classes. The model trained with a dataset with oversampling shows fewer signs of confusion for the minority classes, while remaining at a low validation error. However, balancing oversampling is important since it can lead to overfitting. Here, we choose the degree of oversampling by selecting the parameters that yields both low cross-validation error and small non-diagonal component in the confusion matrix.

**APPENDIX E: TIME EVOLUTION WITH DBSCAN (TIME-DBSCAN) ALGORITHM**

The Time-DBSCAN Algorithm is performed by finding and connecting nearby materials clusters in successive timestamps, which creates historical tracks of materials discovery. First, t-SNE is used to reduce the dimensionality of all neural activations down to two. (We utilized the scikit-learn implementation of t-SNE with parameters "n-component"=2, "perplexity"=30) Entries are then grouped by publication year. For each year, we apply DBSCAN clustering algorithm (scikit-learn, "eps"=0.05, "minimum samples"=2) on the reduced neural activation to form materials clusters. To connect clusters that are close together in this reduced space, we defined a data structure called Link. A Link has a center and population that are used to store and update its location. In the beginning, Links are initialized by materials clusters in the first group by equating their population and center to the Links' population and centers. After initialization, the algorithm runs through groups in each timestamp to update those Links. In each step, we find the materials clusters that are closed to one or more Links in Euclidean distance. To include the nearby materials clusters, the center of each Link is shifted to the population-weighted average of its center and centers of the nearby materials clusters. After that the Link's population is increased accordingly. Materials clusters not included in any existing Link are used to initialize new Links. By repeating this process for each timestamp, the algorithm develops a track of materials systems that share chemical and structural similarity.

**APPENDIX F: LATTICE PARAMETERS PREDICTIONS FOR EACH BRAVAIS LATTICE CLASS**

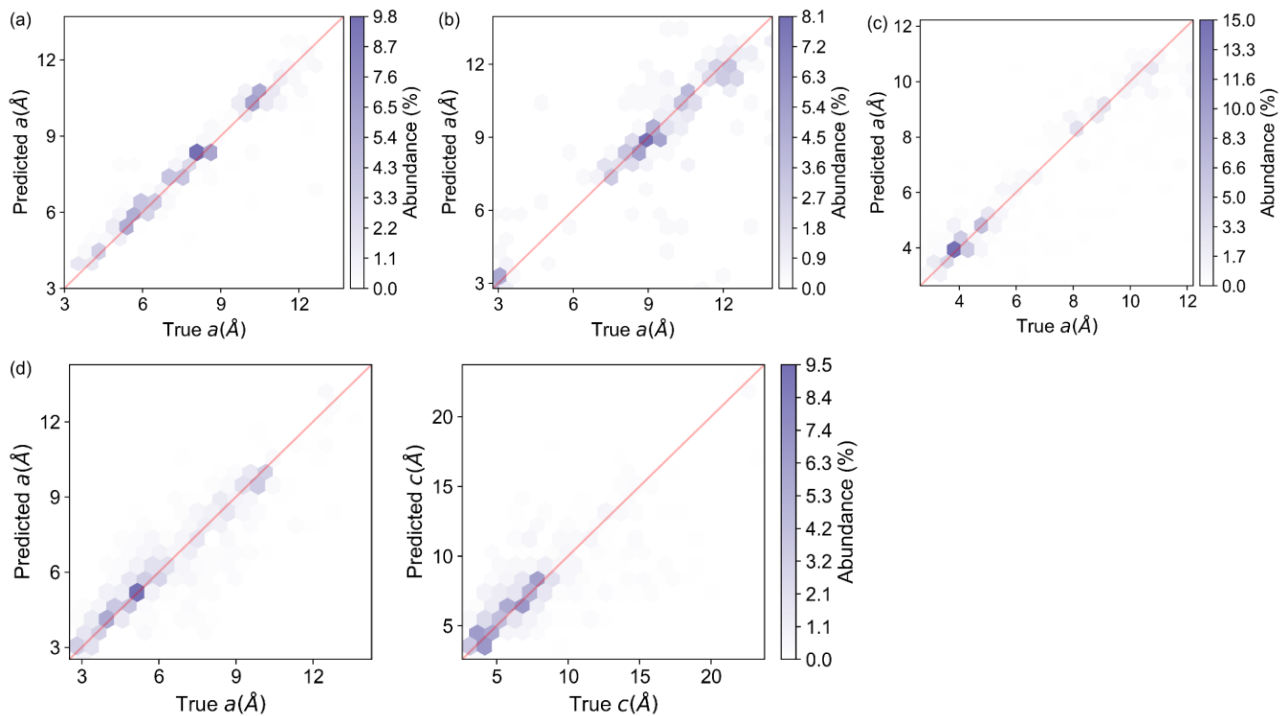

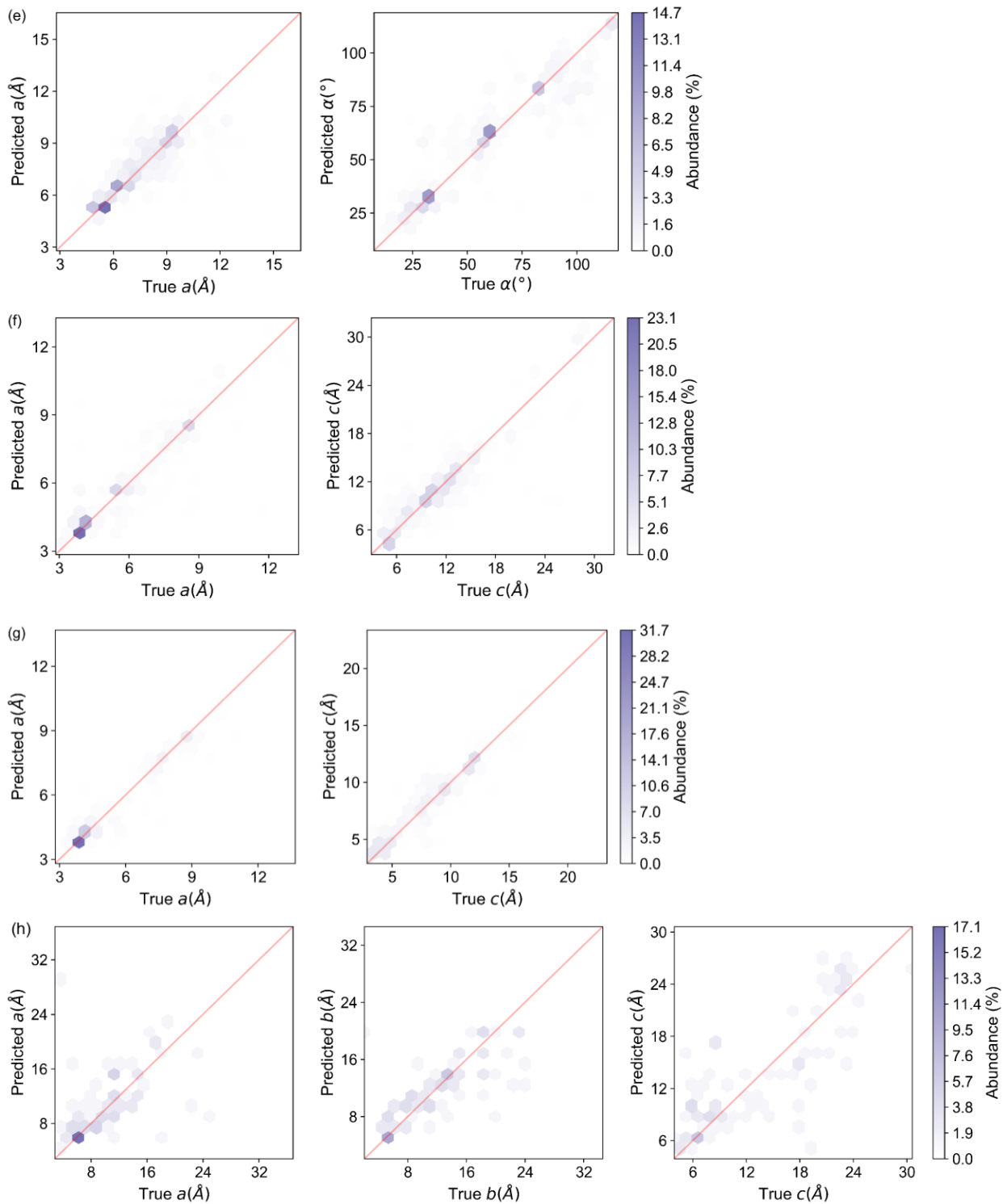

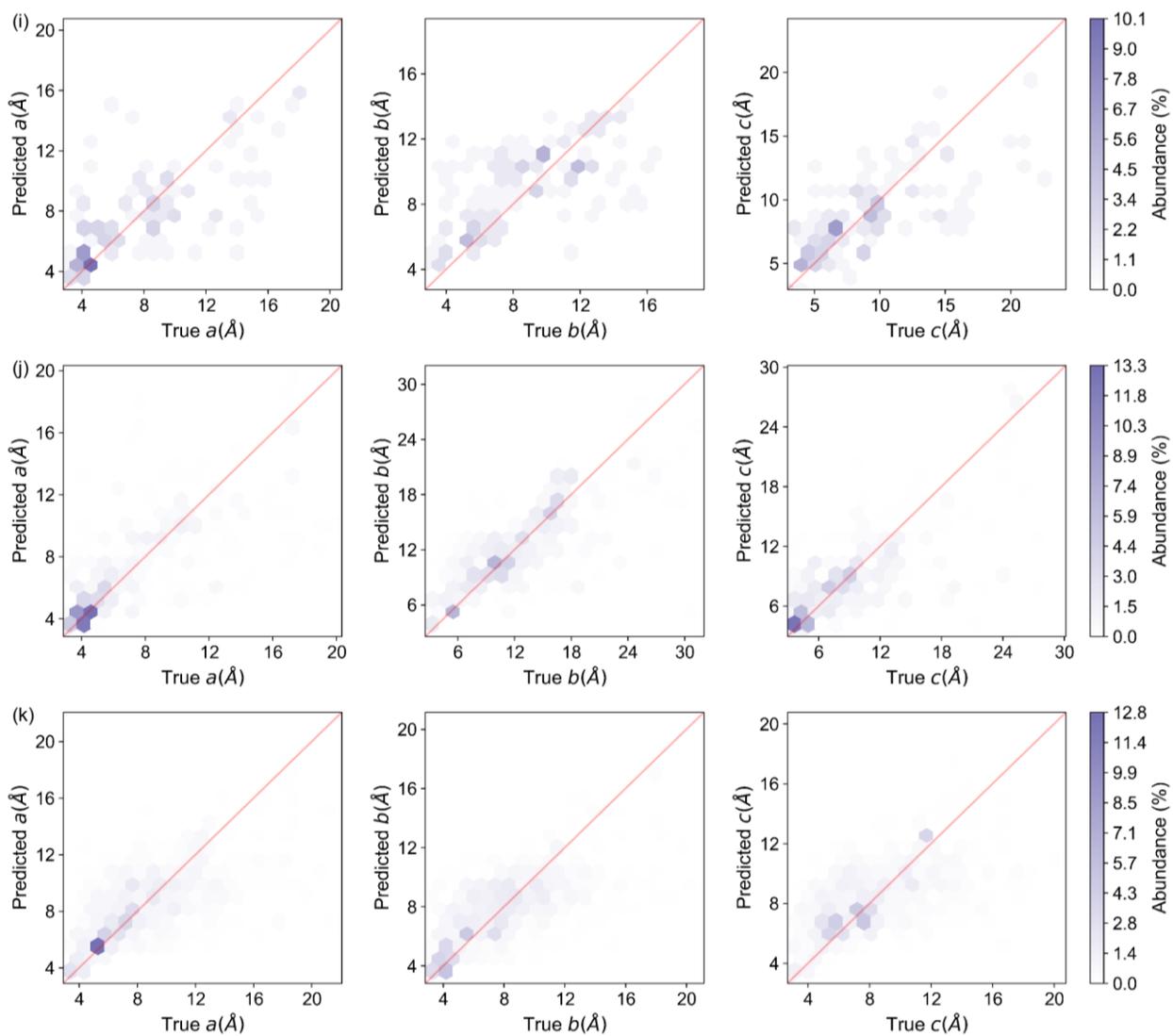

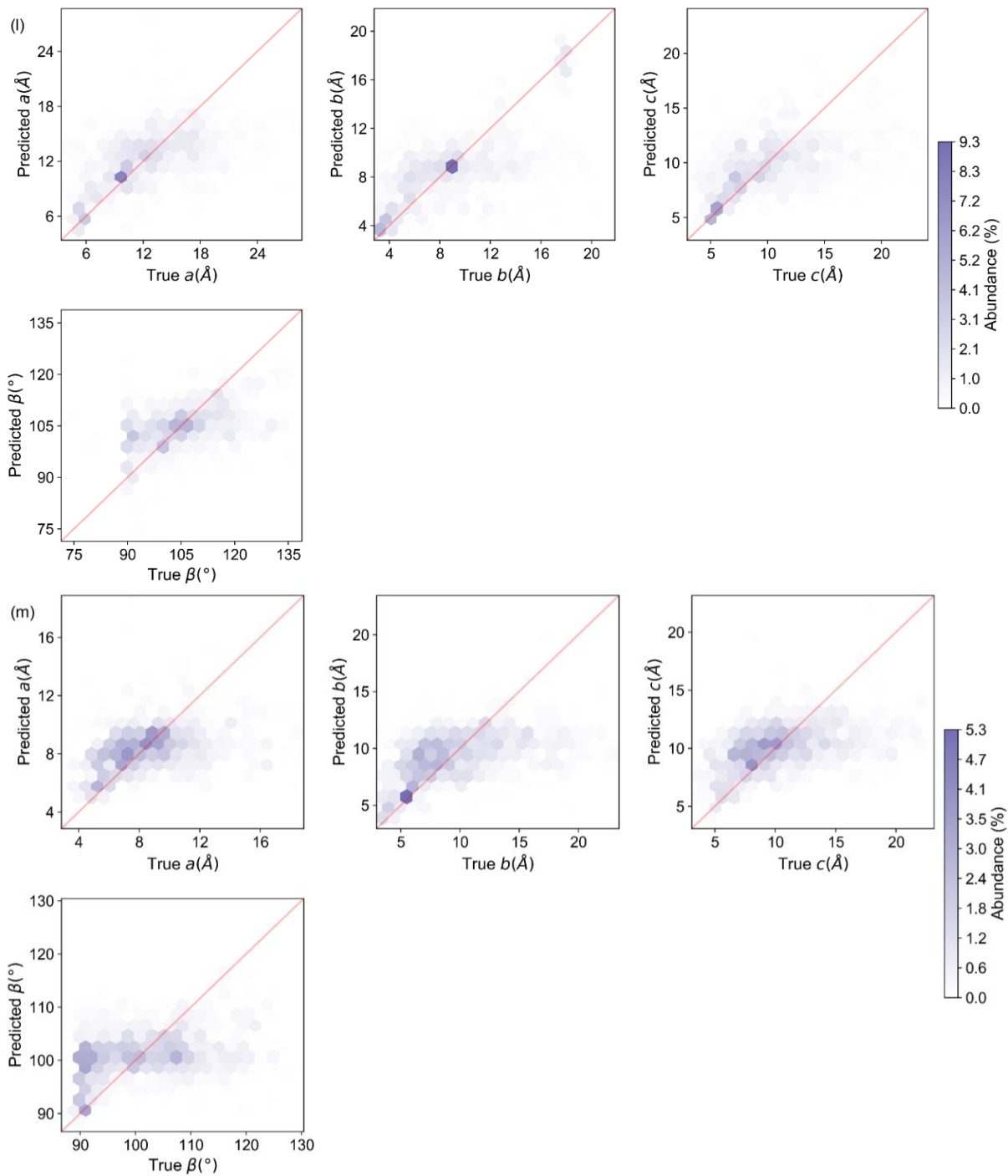

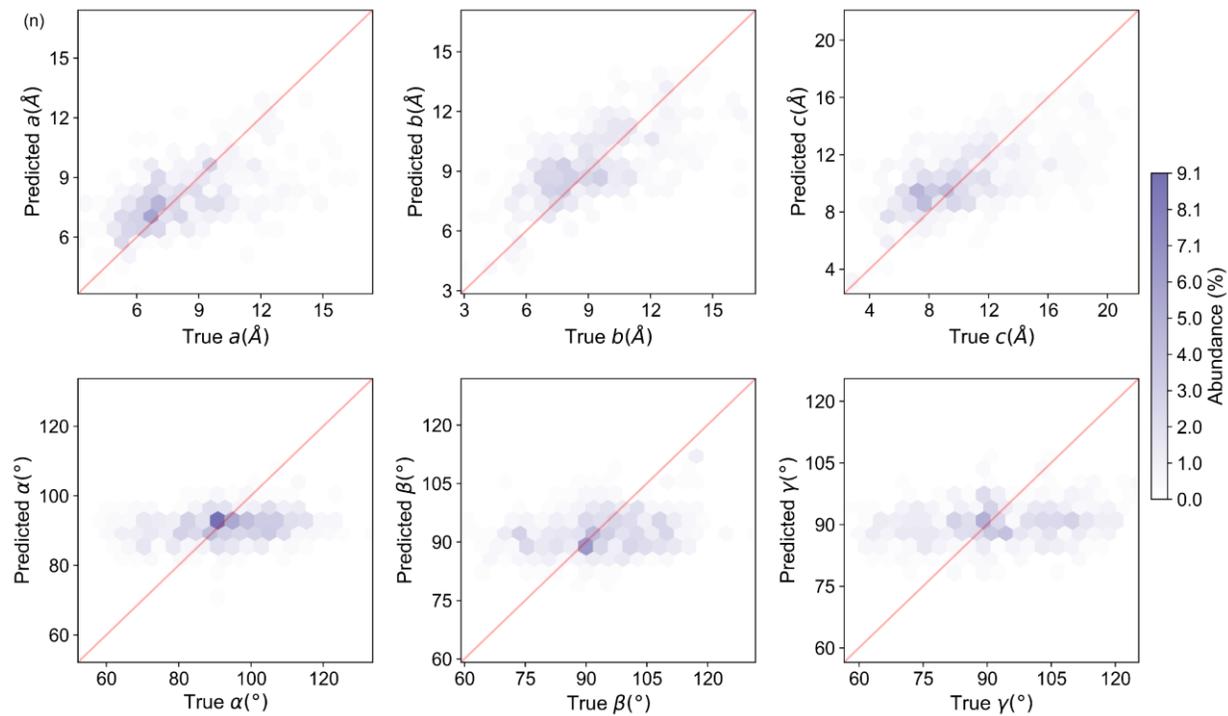

**Fig. A3.** True vs predicted lattice parameters for different Bravais lattices. (a) Cubic (F); (b) Cubic (I); (c) Cubic (P); (d) Hexagonal (P) (e) Rhombohedral (P); (f) Tetragonal (I); (g) Tetragonal (P); (h) Orthorhombic (F); (i) Orthorhombic (I); (j) Orthorhombic (C); (k) Orthorhombic (P); (l) Monoclinic (C); (m) Monoclinic (P); (n) Triclinic (P). The color bar shows the corresponding density of each bin.

**APPENDIX G: EXAMPLES FROM SOME MATERIALS CLUSTERS**

In Table A1 we show examples from several of the materials groups created by clustering the weights of the NN model for the Bravais lattices. We provide the full list for the "11" and "1111" clusters (related to the eponymous iron-pnictide families) as a Supplemental Material. Other lists are available upon request.

Table A1: Compounds included in several clusters formed by the DBSCAN method.

| Compound | Year entered in ICSD | Materials group |
|---|---|---|
| Cuprate Superconductors and related materials | | |
| $YBa_2Cu_3O_{6.6}$ | 1987 | YBCO |
| $ErBa_2Cu_3O_{6.12}$ | 1992 | YBCO |
| $Bi_8Sr_8Mn_4O_{25}$ | 1989 | BSCCO |
| $Bi_2Sr_2YCu_2O_{9.916}$ | 1991 | BSCCO |
| $LaBa_{2.4}Cu_3O_{8.6}$ | 1988 | LBCO |
| $La_{1.8}Sr_{0.2}Co_{0.5}Cu_{0.5}O_{3.95}$ | 1989 | LSCO |
| $HgBa_2Ca_{0.4}Eu_{0.6}Cu_2O_7$ | 1991 | HBCCO |
| $Tl_{1.5}Ba_2Ca_2Cu_{2.1}O_{8.8}$ | 1991 | TBCCO |
| Fe-based Superconductors and related materials | | |
| LiCoAs | 1968 | 111 |
| NaFeAs | 2009 | 111 |
| LaOZnP | 1998 | 1111 |
| $SmOFe_{0.96}Ni_{0.04}As$ | 2010 | 1111 |
| $BaCo_2As_2$ | 1981 | 122 |
| $BaFe_2As_{1.06}P_{0.94}$ | 2010 | 122 |
| $Fe_{1.125}Te$ | 1975 | 11 |
| $FeSe_{0.44}Te_{0.56}$ | 2010 | 11 |
| Solid Li-ion conductors and related materials | | |
| $Na_2Nd_2Ti_3O_{10}$ | 1994 | LLTO |
| $Er_{0.5}Na_{0.5}TiO_3$ | 1998 | LLTO |
| $Li_5La_3Nb_2O_{12}$ | 2006 | Garnet |
| $Li_8La_{18}Fe_5O_{39}$ | 2010 | Garnet |
| $NaTiGeP_3O_{12}$ | 1993 | NASICON |
| $NaNbZrP_3O_{12}$ | 1995 | NASICON |
| $Li_3SO_3N$ | 2013 | LIPON |
| $Li_2PO_4Na$ | 2013 | LIPON |

## APPENDIX H: LEARNING RATE FINDER

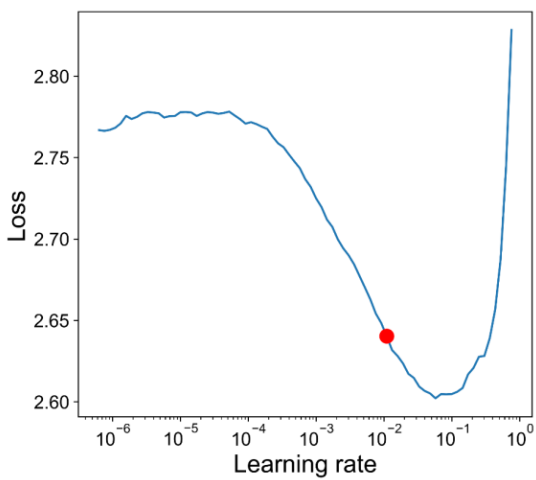

**Fig. A4.** Curve shows minibatch loss change as learning rate changes in each training step. The red dot shows the recommended learning rate.

The basic mechanism of learning rate finder is to determine the largest learning rate that does not destabilize the training. Given an arbitrary NN, the algorithm starts to train the model at a very low learning rate and increments the learning rate after each training step. For each step, the NN uses this learning rate to update its weight. The loss of each step along with the current learning rate is recorded. An example learning rate curve is shown in Fig. A4. A rule-of-thumb is to choose the optimal learning rate with the second order derive that is close to zero. We utilized this method to determine the optimal learning rate of the NN models for Bravais lattice, space group, and lattice constants. The learning rate of $1 \times 10^{-2}$ was found to be a good fit for all models.

## APPENDIX I: CYCLIC LEARNING RATE

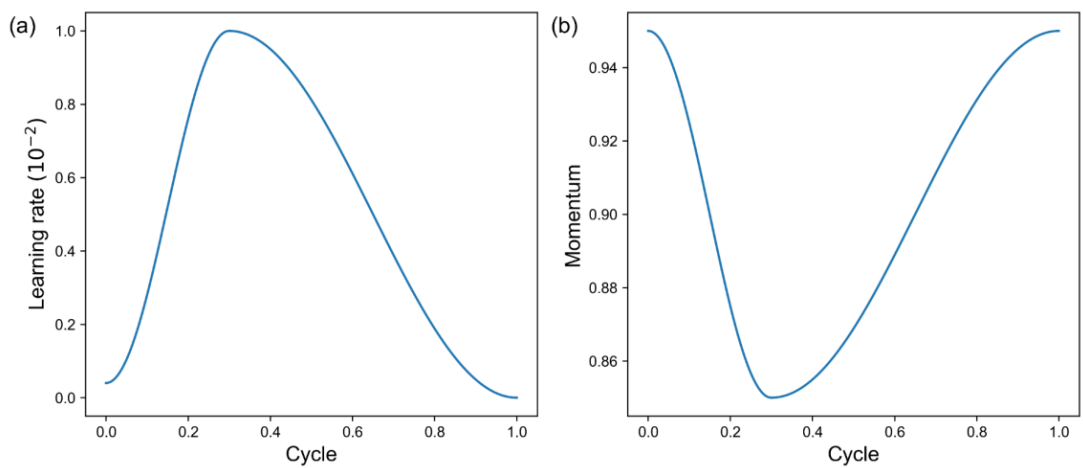

**Fig. A5.** Cyclic learning rate: (a). A typical learning rate schedule as the function of step in each cycle. (b) The adjustment in momentum to accommodate the learning rate change. The cycle length is relative to the actual training iterations, meaning the actual cycle length is a user-defined parameter.

The cycling learning rate policy is an important method to achieve better generalization for less training time[50]. This policy, similar to the learning rate finder methods, varies the learning rate and momentum hyperparameter in each step of the training loop. Fig. A5 shows a typical schedule for one cycle policy where the learning rate increases for half of the cycle and decreases in the other half. The momentum is kept inversely related to learning rate, to balance the change in the learning rate that causes the magnitude of gradient accumulation to increase. The initial learning rate is chosen small, in order to let the Adam optimizer to accumulate a more stable momentum over the gradient. As the momentum become stable, the learning rate is increased to speed up the training process. In the end, the learning rate is gradually reduced to prevent the model from overshooting the correct solution, helping the model to fine tuning to the minimum.